\documentclass[pra,twocolumn,superscriptaddress]{revtex4}

\usepackage{amsfonts}
\usepackage{graphicx}
\usepackage{amsmath}
\usepackage{bm}

\newcommand{\be}{\begin{equation}}
\newcommand{\ee}{\end{equation}}
\newcommand{\bwt}{\begin{widetext}}
\newcommand{\ewt}{\end{widetext}}
\newcommand{\bea}{\begin{eqnarray}}
\newcommand{\eea}{\end{eqnarray}}

\begin{document}
\title{Implementation schemes for unsharp measurements with trapped ions}

\author{Sujit K Choudhary}
\email{choudhary@ukzn.ac.za}
\affiliation{School of Physics, University of KwaZulu, Natal, Durban, South Africa}
\author{T. Konrad}
\email{konradt@ukzn.ac.za}
\affiliation{School of Physics, University of KwaZulu, Natal, Durban, South Africa}
\affiliation{National Institute of Theoretical Physics, University of KwaZulu, Natal, Durban, South Africa}
\author{H. Uys}
\email{huys@csir.co.za}
\affiliation{National Laser Centre, Council for Scientific and Industrial Research, Pretoria, South Africa}

\begin{abstract}
Unsharp POVM measurements allow a variety of measurement applications which minimally disrupt the state
of the quantum system. Experimental schemes are proposed for implementing unsharp measurements on the qubit
levels of a trapped ion.  The schemes rely on introducing weak entanglement between the state of a
target ion, and that of an auxiliary ion, using standard ion trap quantum logic operations, and then realizing an
unsharp measurement through projective measurement on the auxiliary atom.
We analyze common sources of error and their effect on different applications of unsharp measurements.

\end{abstract}

\maketitle

\section{Introduction}

Recent years have witnessed considerable interest in the experimental and theoretical 
study of single quantum systems in atomic traps, optical cavities and coupled quantum dots. 
In trapped ion physics experiments a comprehensive toolbox of control techniques has been developed allowing
implementation of all single and two-qubit quantum gates required for quantum computing \cite{Wineland2003}.  This
quantum control toolbox has enabled new methods of precision measurement \cite{Schmidt2005}, fundamental studies
\cite{Hetet2011} and quantum simulation \cite{Blatt2012}.  A
toolset
which hitherto has not been fully explored experimentally is that of generalized measurements belonging to a
Positive Operator-Valued Measure (POVM) \cite{NielsenChuang}.  The POVM formalism describes all measurements allowable
within the limits of quantum theory.  Such measurements may be implemented by performing a unitary interaction between a
target system and an auxiliary system, and then performing a projective measurement on the auxiliary system.  One
advantage of POVM measurements is that they can be made ``unsharply'' (weakly), so that any individual unasharp
measurment does not project the quantum state into to an eigenstate of the measured observable, and thus does not
strongly disrupt the state dynamics.

The utility of unsharp measurements has
already been demonstrated experimentally in, for example, the quantum feedback stabilization of photon number in a
microwave cavity \cite{Sayrin2011}, and in the improved state preparation of a photon wavefunction through quantum
feedback after a unsharp
measurement \cite{Gillet2010}. We have also shown earlier \cite{Konrad2012}, that it is
possible, in principle, to monitor (approximately) the dynamics (Rabi oscillations) of a single two-level quantum
system in real time using unsharp measurements, or to estimate the Rabi frequency \cite{Audretsch0107}, while
only weakly influencing the original dynamics. By using unsharp measurement estimation in conjunction with feedback,
Vijay et al.~recently demonstrated stabilization of Rabi oscillations in a superconducting qubit at a desired
frequency \cite{VijayMacklin2012}. An experimental set-up for visualizing a photon oscillating between
two cavities has also been proposed \cite{Audretsch0202}. 

This article demonstrates how to implement a class of symmetric POVM measurements on the hyperfine qubit levels of
trapped ions, thus expanding the set of quantum measurement tools in this field.  We exploit standard
techniques of trapped ion quantum control to achieve this goal. The rest of this article is structured as follows.  In
section II we provide some theoretical background on POVM measurement, while in section III we present two schemes for
implementing unsharp measurements.  The efficiency of individual measurements is then analyzed in the light of various
experimental imperfections in section IV.  Finally, in section V we discuss two examples of applications of unsharp
POVM measurements, and the consequences of experimental imperfections on these applications.

\section{POVM Formalism}

Usual quantum measurements project the initial state of a
system to one of the eigenstates of the observables being measured. For example, in a measurement 
of spin along direction $\hat{r}$, the projectors onto the eigenstates are:
\begin{equation}
\hat P_\pm=\frac{1}{2}[\mathbb{I} \pm \hat{r}\cdot\vec{\sigma}],\label{PpPm}
\end{equation}
where $\mathbb{I}$ denotes the identity operator and 
$\vec{\sigma}= (\hat{\sigma_{x}},\hat{\sigma_y}, \hat{\sigma_z})$ is the usual Pauli-operator.  Projective measurements
only represent a restricted set of allowed measurements within quantum theory.
In the more general framework, the states of a
quantum system are represented by a positive trace class operators.
Any observable is represented by a collection of positive
operators $\{E_i\}$ where $0 \le E_i \le \mathbb{I}$ for all $i$ and $\sum_i
E_i = \mathbb{I}$. In a
measurement of such an observable for the state $\rho$ (say), the
probability of occurance of the $i$th result is given by $Tr[\rho
E_i]$. Unlike projective measurement, knowing the operators $\{E_i\}$ is, in general, not enough 
to determine the state of the system after measurement; we further need to know the operators $\hat{M_i}$'s 
constituting the POVM elements $\{E_i\}$. As an example, let $E_i=\displaystyle\sum\limits_{j} M_{ij}^\dagger M_{ij}$,
then after the outcome $i$ is detected, the state is $\rho\mapsto\rho'=\frac{1}{Tr[\rho
E_i]}\displaystyle\sum\limits_{j}M_{ij}\rho M_{ij}^\dagger $. This measurement does not in general preserve the purity
of states, unless there is a single $M_{ij}$ for each $E_i$. In that case the post measured state is given
by 
\begin{equation}
|\psi'\rangle=\frac{M_i|\psi\rangle}{\sqrt{\langle\psi| M_i^\dagger M_i|\psi\rangle}}.\label{stateupdate}
\end{equation}

We will concern ourselves only with the set of symmetric measurement operators:
\begin{eqnarray}\label{unsharp1}
 \hat M_0 &=& \sqrt{p_0}\;\hat P_+ + \sqrt{1-p_0}\;\hat P_-\\
 \label{unsharp2}
\hat M_1 &=& \sqrt{1-p_0}\;\hat P_+ + \sqrt{p_0}\;\hat P_-
\end{eqnarray}
related via
\begin{equation}
M_0^\dagger M_0 +  M_1^\dagger M_1 = \mathbb{I},\label{normeq}
\end{equation}
and $0\leq p_0\leq0.5$. 
The positive operators $M_i^\dagger M_i$, ($i\in{0,1}$) constitute the POVM elements and 
are interpreted as dichotomic observables 
(for example, spin observable of a spin-1/2 particle, energy of a two-level system etc.). The measurement strength
(sharpness) is parametrized by the quantity 
$\Delta p= (1-p_0)-p_0= 1-2p_0$. For $\Delta p=1$, it represents the usual projective (sharp) measurement, while
$\Delta p=0$ represents an infinitely weak (unsharp) measurement, which yields no information and does not change
the state of the system.

\section{Implementation schemes}

We now show that trapped ions provide a realistic  system to implement the generalized measurement
given by Eqs.~(\ref{unsharp1}) and (\ref{unsharp2}). As an example we will choose $\hat
P_\pm=(\mathbb{I}\pm\hat\sigma_z)/2$, but we show in the appendix how to generalize this for arbitrary
projectors Eq.~(\ref{PpPm}). Two schemes are described, both relying on standard
techniques used in ion trap quantum control experiments to generate entanglement.

\subsection{Scheme I}
\begin{figure}
\includegraphics[scale=0.55]{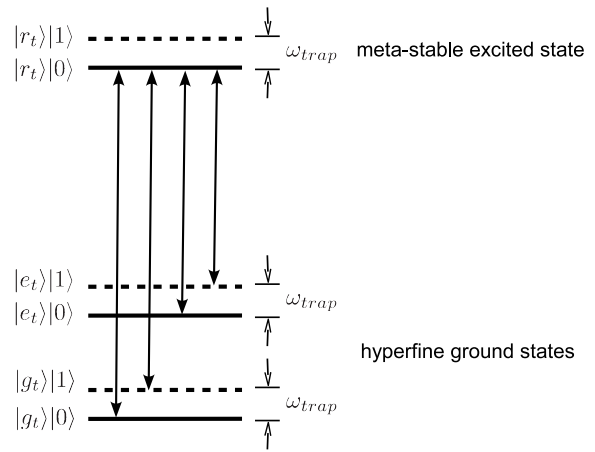}
 \caption{Unsharp measurement of a spin component of a two level transition $|g_t\rangle\longrightarrow|e_t\rangle$
requires an auxiliary
metastable level $|r_t\rangle$ (Scheme I).  In the Lamb-Dicke regime the energy spectrum of the ion exhibits resonance
not only at the carrier laser frequency, but also at the sidebands detuned by a vibrational frequency $\omega_{trap}$. 
The number of excitations of the vibrational mode is indicated by the right-hand ket.}\label{scheme1levels}
\end{figure}

In the first scheme we consider two ions, of different species, to be trapped in the 
same ion trap, thus sharing collective phonon motional modes. 
One ion is the target system, of which we wish to measure some observable unsharply.  We will label its internal
state with the ket $|\psi_t\rangle$. The second is an auxiliary ion, described by $|\psi_a\rangle$, which is used both
for sympathetic
cooling of the phonon modes, and for assisting in implementing the unsharp measurement.  Since the ions are of
different species, any light field used to manipulate one ion does not affect the other ion, obviating the need
for independent ion addressing.  

Both ion species need to be carefully chosen to have appropriate level structures to enable the measurement scheme. 
For the target ion we choose two hyper-fine (qubit) ground levels $|g_t\rangle$ and $|e_t\rangle$ and aim to measure the
$z$-component of
its effective spin, $\hat\sigma_z$, unsharply. This is accomplished with the assistance of a third meta-stable
excited state, $|r_t\rangle$, as shown in Fig.~\ref{scheme1levels} (or a stimulated Raman transition to a third
hyperfine groundstate). A meta-stable state is chosen so as to minimize the effects of
spontaneous emission.   
Figure \ref{scheme1levels} further indicates that the ion's energy spectrum incorporates vibrational excitations of the
trap, i.e.
$|g_t\rangle|0\rangle$ describes the state where the ion is in the electronic ground state, $|g_t\rangle$, and the
vibrational state, $|0\rangle$,
of the ion has zero phonons (vibrational ground state).
The internal structure of the auxiliary atom must allow Quantum Logic
Spectroscopy (QLS) \cite{Schmidt2005} measurements as will be discussed later.

The measurement scheme can be divided into four stages:\\ 
(1) Sympathetic cooling to the vibrational ground-state  on the
auxiliary ion.\\
(2) Repump state preparation of the auxiliary atom.\\
(3) Unitary unsharp measurement preparation of both ions.\\
(4) Unsharp measurement execution through QLS on the auxiliary ion.\\
Moreover, the system is assumed to be in the Lamb-Dicke regime \cite{Wineland1998} so that the vibrational
side-bands of the internal transitions are resolved. This implies that a light field detuned to the red (blue) of a
ground to excited-state transition by a motional mode frequency $\omega_{trap}$, will lead to the absorption (emission)
of a vibrational phonon, and \textit{vice versa}. 

\begin{figure}
\includegraphics[scale=0.45]{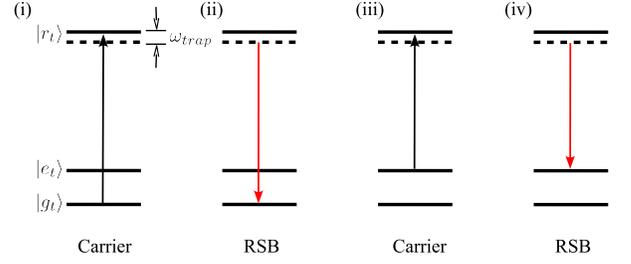}
 \caption{(Color online) Sequence of laser-pulses on an ion to prepare it to the desired state Eq.~(\ref{5}) for unsharp
measurement of $\hat\sigma_z$ through Scheme I.}\label{scheme1seq}
\end{figure}

If initially the target atom is prepared in a coherent superposition of its qubit levels $|\psi_t^{(0)}\rangle =
c_1|g_t\rangle+c_2|e_t\rangle$, then,  after the ground-state
cooling stage on the auxiliary ion, the state of the system can be described by:

\begin{eqnarray}\label{1}
 |\psi\rangle&=&|\psi_t^{(0)}\rangle|\psi_a\rangle|0\rangle\\
             &=&(c_1|g_t\rangle+c_2|e_t\rangle)|\psi_a\rangle|0\rangle,
\end{eqnarray}
where $|c_1|^2+|c_2|^2=1$. The measurement preparation stage consists of a sequence of four laser pulses applied to the
target atom (as illustrated in Fig.~\ref{scheme1seq}):\\
(1) A pulse on-resonance with the transition $|g_t\rangle\longrightarrow|r_t\rangle$ (carrier-pulse) for a duration
$t=(2/\Omega_1)\cos^{-1}(\sqrt{p_0})$ (Fig.~\ref{scheme1seq}(i)),  
($\Omega_1$ is the Rabi-frequency connecting $|g_t\rangle$ and $|r_t\rangle$),
leading to the state:
\begin{equation}\label{2}
 |\psi_1\rangle=\left[c_1(\sqrt{p_0}|g_t\rangle+\sqrt{1-p_0}~|r_t\rangle)+
c_2|e_t\rangle\right]|\psi_a\rangle|0\rangle
\end{equation}
(2) A pulse on the same transition, but red-detuned by a motional mode frequency (red side-band, RSB, pulse
(Fig.~\ref{scheme1seq}(ii)),
causing the transition $|r_t\rangle |0\rangle\longrightarrow|g_t\rangle|1\rangle$. The component
$|g_t\rangle|0\rangle$ of $|\psi_1\rangle$ remains unaffected by this pulse as the state 
$|r_t\rangle|\!-\!\!1\rangle$ does not exist \cite{Wineland2003, Wineland1998}. 
This results in the following two ion state:
\begin{equation}\label{3}
 |\psi_2\rangle=\left[c_1|g_t\rangle(\sqrt{p_0}|0\rangle+\sqrt{1-p_0}~|1\rangle)+
c_2|e_t\rangle|0\rangle\right]|\psi_a\rangle
\end{equation}
(3) A carrier pulse on resonance with the transition $|e_t\rangle\longrightarrow|r_t\rangle$ of duration
$t=(2/\Omega_2)\cos^{-1}(\sqrt{1-p_0})$, 
($\Omega_2$ is the Rabi-frequency connecting $|e_t\rangle$ and $|r_t\rangle$), (Fig.~\ref{scheme1seq}(iii)). This
transforms the state
to
\begin{align}\label{4}
 |\psi_3\rangle= \left[c_1|g_t\rangle(\sqrt{p_0}|0\rangle+\sqrt{1-p_0}~|1\rangle)+\right.\nonumber\\
 \left.c_2(\sqrt{1-p_0}|e_t\rangle+\sqrt{p_0}~|r_t\rangle)|0\rangle\right]|\psi_a\rangle
\end{align}
(4) Finally, a red side-band pulse between $|e_t\rangle$ and $|r_t\rangle$ (Fig.~\ref{scheme1seq}(iv)) yielding
\begin{align}\label{5}
|\psi_4\rangle=&\left[c_1|g_t\rangle(\sqrt{p_0}|0\rangle+\sqrt{1-p_0}|1\rangle)\right.\nonumber\\ 
& \qquad\left.+\;c_2|e_t\rangle(\sqrt{1-p_0}|0\rangle+\sqrt{p_0}|1\rangle)\right]|\psi_a\rangle\nonumber\\
=&\left[(\sqrt{p_0}~c_1|g_t\rangle+ \sqrt{1-p_0}~c_2|e_t\rangle)|0\rangle\;+\right.\nonumber\\
&\qquad\left.(\sqrt{1-p_0}~c_1|g_t\rangle+ \sqrt{p_0}~c_2|e_t\rangle)|1\rangle\right]|\psi_a\rangle.
\end{align}

The generalized measurements Eqs.~(\ref{unsharp1}) and (\ref{unsharp2}) on the internal degrees of freedom of the 
target ion can now be realized by 
performing a projective measurement on the vibrational state, because measuring in the $\{|0\rangle,|1\rangle\}$ basis 
projects the internal state of the target ion onto either of the two states:
\begin{align}
&(\sqrt{p_0}~c_1|g_t\rangle+ \sqrt{1-p_0}~c_2|e_t\rangle)\nonumber\\
&(\sqrt{1-p_0}~c_1|g_t\rangle+ \sqrt{p_0}~c_2|e_t\rangle).
\end{align}
This sequence is equivalent to applying the measurement operators $\hat M_0$ or $\hat M_1$, to the initial state of
the target atom:
\begin{eqnarray}\nonumber
 \hat M_0|\psi_t^{(0)}\rangle; \hat M_0=\sqrt{p_0}~|g_t\rangle\langle g_t|+\sqrt{1-p_0}~|e_t\rangle\langle e_t|\\
\label{unsharp4}
\hat M_1|\psi_t^{(0)}\rangle; \hat M_1=\sqrt{1-p_0}~|g_t\rangle\langle g_t|+\sqrt{p_0}~|e_t\rangle\langle e_t|.
\end{eqnarray}
\begin{figure}
\includegraphics[scale=0.65]{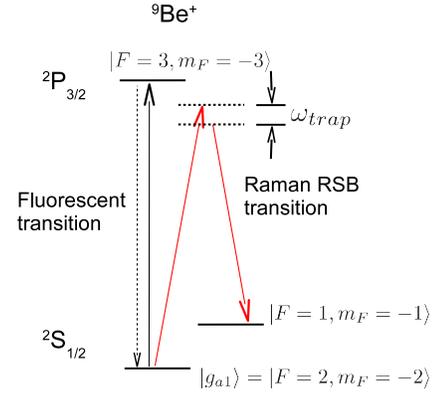}
\caption{(Color online) Quantum logic spectroscopy detection of the motional state of the two-ion
system. A Raman red
sideband pulse, red (grey) lines, transfers the auxiliary ion to a dark state, $|^{2}{S}_{\frac{1}{2}}, F=1,
m_F=-1\rangle$, only if a vibrational excitation is present.
Subsequent fluorescence (or not) on the $|^{2}{S}_{\frac{1}{2}}, F=2, m_F=2\rangle\longrightarrow
|^{2}{P}_{\frac{3}{2}}, F=3, m_F=3\rangle$ transition projects the vibrational state onto $|0\rangle$
$(|1\rangle)$ \cite{Schmidt2005}.}\label{QLSion}
\end{figure}
Projection onto the vibrational state is accomplished through a QLS measurement \cite{Schmidt2005} which
we briefly review.  QLS maps the vibrational state of the two ion system onto a dark (bright) hyperfine, ground state
level of the auxiliary ion, conditioned on the presence (or not) of a vibrational excitation.  A subsequent
fluorescence detection measurement of the auxiliary ion state completes the measurement.  Let's denote the qubit
levels of the auxiliary ion with $|\!\downarrow\rangle, |\!\uparrow\rangle$. The required level structure
for the auxiliary ion is shown in Fig.~\ref{QLSion}.  $^9$Be$^+$ is a suitable candidate if we associate:
$|\!\downarrow\rangle\equiv |^{2}{S}_{\frac{1}{2}}, F=2, m_F=-2\rangle$ and $|\!\uparrow\rangle\equiv
|^{2}{S}_{\frac{1}{2}},
F=1, m_F=-1\rangle$. We assume that the internal state of the auxiliary ion was initially prepared in level
$|\!\downarrow\rangle$, so that after the measurement preparation sequence the full stat of the system is:  
\begin{align}\label{6}
|\Psi\rangle&=[\sqrt{p_0}~c_1|g_t\rangle+ \sqrt{1-p_0}~c_2|e_t\rangle]|0\rangle|\!\downarrow\rangle\nonumber\\
&+[\sqrt{1-p_0}~c_1|g_t\rangle+ \sqrt{p_0}~c_2|e_t\rangle]|1\rangle|\!\downarrow\rangle.
\end{align}
A Raman red side-band pulse on the auxiliary ion, illustrated in Fig.~(\ref{QLSion}),  transfers the final state of 
the  two-ion system to
\begin{align}
|\Psi_f\rangle&=\left[(\sqrt{p_0}~c_1|g_t\rangle+ \sqrt{1-p_0}~c_2|e_t\rangle)|\!\downarrow\rangle\right.\nonumber\\
&\left.+(\sqrt{1-p_0}~c_1|g_t\rangle+ \sqrt{p_0}~c_2|e_t\rangle)|\!\uparrow\rangle\right]|0\rangle.\label{finalpsi}
\end{align}
Because $|\!\downarrow\rangle$ fluoresces strongly, whereas $|\!\uparrow\rangle$ is dark to the detection light
\cite{Schmidt2005}, we can measure the final state of the auxiliary ion sharply, thereby executing the unsharp
measurement, Eq.~(\ref{unsharp4}), on the target ion.

\begin{figure}
\includegraphics[scale=0.55]{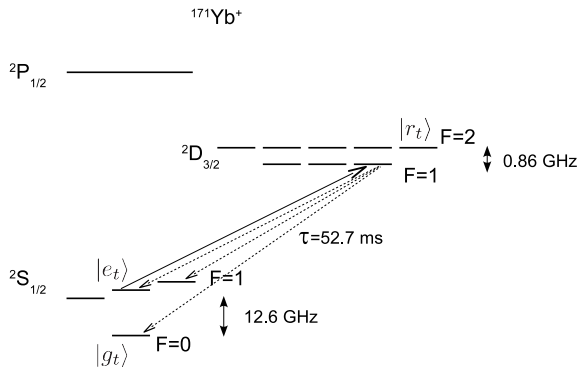}
 \caption{Energy levels of an $^{171}\textnormal{Yb}^{+}$ ion. The hyperfine ground states
$|^{2}{S}_{\frac{1}{2}}, F=0, m_F=0\rangle$ and $|^{2}{S}_{\frac{1}{2}}, F=1, m_F=0\rangle$ (clock states) serve as
qubit levels undergoing Rabi oscillations. The metastable state $|^{2}{D}_{\frac{3}{2}}, F=1, m_F=+1\rangle$ assists
in implementing the unsharp measurement.  The dashed arrows show possible spontaneous decay paths. Raman scattering
 to the $|^{2}{S}_{\frac{1}{2}}, F=1, m_F=+1\rangle$ state removes the
ion from the oscillating qubit transition and halts the experiment. \label{yblevels}} 
\end{figure}

\subsection{Scheme II}
Scheme I discussed above has the advantage that two different ion species are used, making the target qubit states
immune to spontaneous emission from laser light addressing the auxiliary ion.  We now discuss a different scheme using
two ions of the same species, but with a simpler pulse sequence.  The scheme is
based on the Kitegawa squeezing Hamiltonian \cite{Kitagawa1993}.  
\begin{equation}\label{Kitegawa}
 H_S=\hbar\frac{\chi}{2} J_z^2,
\end{equation}
where $\hat J_z = \sum\limits_j\hat\sigma^z_j$ is the collective spin operator.  This Hamiltonian is routinely used to
implement geometric phase gates in quantum information processing experiments, by exerting a spin-state dependent
optical
dipole force on the ions using far off-resonant laser beams \cite{Leibfried2003}. Since the experimental approach is
well known, we will simply discuss how it enables unsharp measurement  

We again consider two ions, target and auxiliary, in the same trap.   The ions have suitable qubit levels and excited
states allowing repumping, cooling and fluorescence detection. The scheme consists of four stages:\\ 
(1) Cooling of the auxiliary ion, thus cooling the target ion sympathetically.\\
(2) State preparation (repump) of the auxiliary ion\\
(3) Unsharp measurement preparation on both ions\\
(4) Fluorescence detection of the auxiliary ion state.\\
Since both ions are of the same species, single ion addressing with tightly focused laser beams is required during the
cooling and fluorescence detection stages, to prevent the light manipulating the auxiliary ion from destroying the
coherence in the target ion.  Alternatively, a third ion of a different species can be used for sympathetic cooling
purposes.

Let's denote Pauli operators acting on the target state by $\hat\sigma_j$, and Pauli operators acting on the auxiliary
atom by $\overline\sigma_j$, for $j=x,y,z$.  Then, after cooling and optical pumping of the auxiliary ion into the upper
qubit level, step (3) of the scheme can be represented by the following time evolution operator
\begin{equation}
 \hat U = e^{i\frac{\pi}{4}\overline\sigma_y}e^{i\chi\hat\sigma_z\overline\sigma_z}e^{i\frac{\pi}{4}\overline\sigma_x}.
\end{equation}
The single particle operations on the auxiliary ion can be carried out with a stimulated Raman interaction acting only
on that ion. It is easy to show that choosing $\chi\ll1$ leads to a final state
\begin{eqnarray}
 |\psi_f\rangle &=&e^{i\frac{\pi}{4}}\left\{\left[ic_1(1+\chi)|g\rangle +
ic_2(1-\chi)|e\rangle\right]|\!\downarrow\rangle\right.\\
&+&\left. \left[c_1(1-\chi)|g\rangle +c_2(1+\chi)|e\rangle\right]|\!\uparrow\rangle\right\}.
\end{eqnarray}
Now making a projective measurement of the auxiliary atom spin-state, step (4), realizes the unsharp measurement on the
target state, similarly to the final step in scheme I.  Note that the effect of the squeezing operator
$\chi\hat\sigma_z\overline\sigma_z$ is equivalent to that of the Kitegawa operator $(\chi/2) \hat J_z^2$ up to a global
phase factor that can be neglected.

\section{Unsharp measurement fidelity}

Each of the schemes presented above is subject to experimental imperfections.  We now take stock of the most important
sources of errors that would deteriorate the efficacy with which each unsharp measurement can be executed.  In what
follows we will assume throughout that the laser light interacting with the ion, has 5 mW of power focussed to a spot
size with a diameter of 50 $\mu m$, a conservative estimate for the light intensities achievable in a typical
laboratory at UV wavelengths.

\subsection{Scheme I fidelity}

For concretenes we take $^{171}$Yb$^+$ as a candidate for the
target atom in the scheme I implementation.  The clock transition, $|^{2}{S}_{\frac{1}{2}}, F=0,
m_F=0\rangle\longrightarrow|^{2}{S}_{\frac{1}{2}},
F=1,m_F=0\rangle$ is chosen as a qubit. In addition a suitable metastable level is
needed for implementation of this scheme.  The $|^{2}{D}_{\frac{3}{2}}, F=1, m_F=+1\rangle$ with a life time of
$\tau=52.7$
ms can serve this purpose. The level scheme for $^{171}$Yb$^+$ is illustrated in Fig.~\ref{yblevels}.

\subsubsection{Spontaneous emission}
While the $^2$D$_{3/2}$ state, $|r_t\rangle$, in ytterbium is metastable (electric quadrupole transition), its
$\tau_{sp}=52.7$ ms lifetime does not make the scheme completely immune to spontaneous emission.  To find an upper
bound to the effect of spontaneous emission we calculate the emission rate if an ion would spend the entire time
it takes to implement the four preparation pulses in $|r_t\rangle$.  The time $\delta t$ to implement the four pulses is
roughly set by the RSB pulse rather than the carrier pulse, since for the
RSB pulse the coupling with the laser field is reduced by a factor corresponding to the Lamb-Dicke parameter,
$\eta_{LD}$ \cite{Wineland1998}, as compared to the carrier pulse. The lifetime of the metastable state allows us to
estimate the quadrupole matrix element from
\begin{equation}
 \mu_q = \sqrt{\frac{3\pi c\epsilon_0\gamma_q}{\omega^3}}
\end{equation}
Where $\gamma_q=1/\tau_{sp}$, $c$ is the speed of light, $\epsilon_0$ the permittivity of free space, and $\omega_q$
the angular frequency of the transition.  For our typical laser parameters (5 mW and a focus spot width of diameter 50
$\mu$m), we find an electric field strength estimate of $E_0=\sqrt{2I/(c\epsilon_0)}\approx22$ kV/m.  This
predicts a RSB pulse coupling frequency of $\Omega_{RSB}=2\pi/\tau_{RSB}=\eta_{LD}\mu_qE_0/(4\hbar)\approx
2\pi\times(70 \textnormal{ kHz})$.
The probability that the ion did \textit{not} spontaneously emit from the level $|r_t\rangle$ at the end of
$\delta t\approx2\tau_{RSB}\approx15\;\mu$s is $\exp{(-\delta t/\tau_{sp})}$. We can therefore calculate the
number of excitations to $|r_t\rangle$ which predict a 50\% chance of
decay as $N=-(\tau_{sp}/\delta t)\ln(0.5)\approx1100$, or otherwise stated, the probability for
spontaneous scattering per measurement is $P_{sp}=0.0007$.

It is worth noting that different spontaneous emission channels exist, not all being equally deleterious.  The ion might
Rayleigh scatter back into the ground state from which it is driven, Raman scatter into the other qubit state, or
even Raman scatter out of the qubit transition.
A Raman scattering event to a different qubit state will completly destroy the coherence in the target qubit, and
situations can arise in which Rayleigh scattering also becomes strongly decohering \cite{Uys2010}.  

Spontaneous emission problems can be overcome by choosing an
ion with a much longer lived  metastable state, or with no spontaneous emission path that would remove the ion from the
qubit levels.

\subsubsection{Imperfect Quantum Logic Spectroscopy}
The state mapping during QLS is not perfect. Incorrect mapping might, for example,
occur because external heating of the ion causes an unintended excitation of the motional mode \cite{Deslauriers2006},
or due to a spontaneous scattering from the auxiliary ion. During an unsharp measurement sequence this
translates to the
experimenter measuring a false outcome $\hat M_0$ instead of $\hat M_1$ or \textit{vice versa}.  Since external factors
such as motional heating is very much dependent on the specific ion trap geometry, of which there are many kinds, we
will not attempt to estimate the typical probability of incorrect mapping.  Instead we will only evaluate
experimental fidelities for different applications of unsharp measurements, at different probabilities of incorrect
mapping, in later sections.

\subsection{Scheme II fidelity}

Measurement fidelity in our second scheme will also be be limited by the effects of spontaneous emission.  The detection
scheme, however, does not rely on QLS and the concomitant state mapping errors are absent.  

\subsubsection{Spontaneous emission from optical dipole fields}
Whereas in scheme I we were concerned with decoherence due to spontaneous emission by an atom occupying a
metastable excited state, in scheme II  we consider the spontaneous scattering of far \textit{off-resonant}
light fields.  This is a well-studied problem \cite{Ozeri2005, Uys2010}, and depends sensitively on the polarizations of
the light fields, frequency detunings and detailed atomic structure. Here we make simple order of magnitude estimates
following a typical scheme for a squeezing Hamiltonian as used in \cite{Leibfried2003} to implement
geometric phase gates. We are interested in finding the time needed to implement weak squeezing using this scheme,
 and compare it to the spontaneous emission expected during that time.
 As in \cite{Leibfried2003} we will consider two beryllium ions in a trap with stretch-mode
frequency of $\omega_s\approx(2\pi)\times6$ MHz. This motional mode is excited off-resonantly by two interfering
laser beams impinging on the trapped ions.  The excitation is executed in such a way that the
vibrational wavepacket traces out a closed loop in phase space.
A phase gate is fully implemented when the geometric phase, $\phi_G$,
which the ions have accumulated is equal to $\pi/2$ \cite{Ozeri2011}.
This phase is related to the optical dipole force, $F_0$,  via
\begin{equation}
\phi_G = \frac{\pi}{2}\left(\frac{F_0z_0\tau_g}{\hbar}\right)^2,\label{geophi}
\end{equation}
here $z_0=\sqrt{\hbar/(2m\omega_s)}$ is the width of the stretch-mode ground state, and $m$ the mass of the ion.  The
laser beams enter at
90$^{\textnormal{o}}$ with respect to each
other, forming an interference pattern with an effective wave number $k_{eff}=2\pi\sqrt{2}/(313 \textnormal{nm})$. 
Assuming a realistic Lamb-Dicke parameter of $\eta_{LD}=k_{eff}z_0\approx0.2$ implies an associated width $z_0\approx7$
nm. 
The lasers fields cause an AC-Stark energy shift, $E_S$, on the atomic transitions, with gradients due to the
interference pattern on the order of  $\partial 
E_S/\partial z = \mu^2E_0^2k_{eff}/4\hbar^2\Delta$, where $\Delta$ is the detuning of the lasers from the excited state
($^2$P$_{1/2}$) transition. Sticking with typical laser parameters 5 mW and 50 micron focus diameter, the
gradient predicts an optical dipole force of $F_0=\hbar\partial E_S/\partial z\approx 16$ zeptonewton
(zepto$=10^{-21}$). The
corresponding time to complete a phase gate is about $\tau_g\approx3\;\mu$s.

 The on-resonance coupling strength to the excited state is $g=\mu E_0/\hbar$, but because the laser beam is far off
resonance there exists only a very small probability to occupy the excited state, of order $P_u\approx g^2/\Delta^2
\approx2\times10^{-5}$.  During an unsharp measurement only weak squeezing will be implemented, say for a time
$\tau_g/5$, which is about $n_{sp}\approx23$ spontaneous decay lifetimes.  The probability for a spontaneous emission
event over this entire time is roughly $P_{sp}=1-(1-P_{u})^{n_{sp}}\approx0.0005$.  

Given the probability $P_{sp}$ about 1400 unsharp measurements can be carried out before the likelihood for no
spontaneous emission event has decayed to 50\%.

\section{Applications of unsharp measurements}

 In many applications of POVM measurements a sequence of successive measurements are required.  Having investigated the
imperfections in single measurements for each of our proposed schemes, we now look at the cumulative effect of many
measurements in specific applications.  In particular we discuss: (1) The efficacy with which real-time state estimation
can be done of a two level system undergoing Rabi oscillations, (2) The efficiency with which qubit states can be
prepared using only POVM measurements following a scheme proposed by \cite{Ashab2010}.

\subsection{State estimation fidelity}

In previous work \cite{Konrad2012} we've demonstrated that it is in principle possible to monitor quantum dynamics (Rabi
oscillations) by executing a sequence of periodic, unsharp measurements as the quantum system evolves.   In the presence
of classical noise fields this state estimation is still possible, albeit with reduced fidelity. For this estimation
process to be successfull the relevant time scales in the experiment must obey the inequality
condition
\begin{equation}
 \tau_{meas}\ll\tau_R\ll\tau_{m}\ll\tau_N.\label{conditions}
\end{equation}
Here $\tau_{meas}$ is the duration of a single unsharp measurement, $\tau_R$ is the Rabi frequency,
and $\tau_N$ is the decoherence time due to the external noise processes.  $\tau_m$ is the expected time it
takes an arbitrary initial state to reduce to an eigenstate of the measured observable in the absence of dynamics other
than measurement \cite{AudretschDiosiKonrad02}.

We now calculate the dependence of estimation fidelity on a number of noise
sources and operational errors.  These include:\\
(1) Classical dephasing noise resulting from magnetic field fluctuations, fluctuating Stark shifts and reference
oscillator noise.\\
(2) Imperfect state mapping during QLS.\\
(3) Spontaneous emission.\\
We now discuss each noise source in turn.

\subsubsection{Classical dephasing noise}
Let's establish whether experimental parameters which will allow us to fulfull condition Eq.~(\ref{conditions})
are realizable.  Classical dephasing noise can be described by a Hamiltonian $H=\beta(t)\hat\sigma_z$. One convenient
measure of the strength of
the noise is the ratio of the root-mean-square noise stength, $\Delta\beta$, to the Rabi frequency.  This ratio can be
determined if the Ramsey coherence time of the qubit states is known. 
The clock transition is insensitive to magnetic field fluctuations leading to long 
coherence times.  In ytterbium this transition has been reported to have a Ramsey coherence time of 2.5
seconds \cite{Olmschenk2007}. 
The Ramsey coherence is related to the noise fieldstrength
through the noise power spectrum via the expression \cite{Uhrig2008}
\begin{eqnarray}
 s(t)&&=e^{-\frac{8}{\pi}\int^\infty_0\frac{P(\omega)}{\omega^2}\sin{^2(\omega t/2)}d\omega}\\
&&\approx e^{-\frac{2}{\pi}\int^\infty_0P(\omega)d\omega t^2},\label{Ramscoh}
\end{eqnarray}
where $s(t)$ is the coherence, $P(\omega)$ is the power spectrum of the noise field $\beta(t)$, and the last line
follows when $\tau\ll1$. We
further know that $\Delta\beta^2 = \frac{1}{\pi}\int^\infty_0P(\omega)d\omega$, which by comparing to
Eq.~(\ref{Ramscoh}) allows us to relate the Ramsey coherence time to the mean field strength as
$\tau_{Ramsey}=\tau_N=1/(\sqrt{2}\Delta\beta$).
 Experimentally achievable Rabi oscillation time can be much shorter than $\tau_{Ramsey}$, e.g.~around 12 $\mu$s
reported in \cite{Olmschenk2007}. However, to fulfill condition Eq.~(\ref{conditions}) one can experimentally choose
$\tau_R\approx100$
ms. 
Then
with $\tau_{Ramsey}=2.5$ s one finds $\Omega_R/\sqrt{2}\Delta\beta=\tau_N/\tau_R\approx200$.  Given perfect,
instantaneous unsharp
measurements, numerical simulations have shown that this will predict estimation fidelities in excess of $F=0.999$
\cite{Konrad2012}. However, several further imperfections need to be considered.  We also note that oscillator noise can
at any time
be introduced artificially into the system to enable testing of the system.

\subsubsection{Imperfect State Mapping}

\begin{figure}
\includegraphics[scale=0.45]{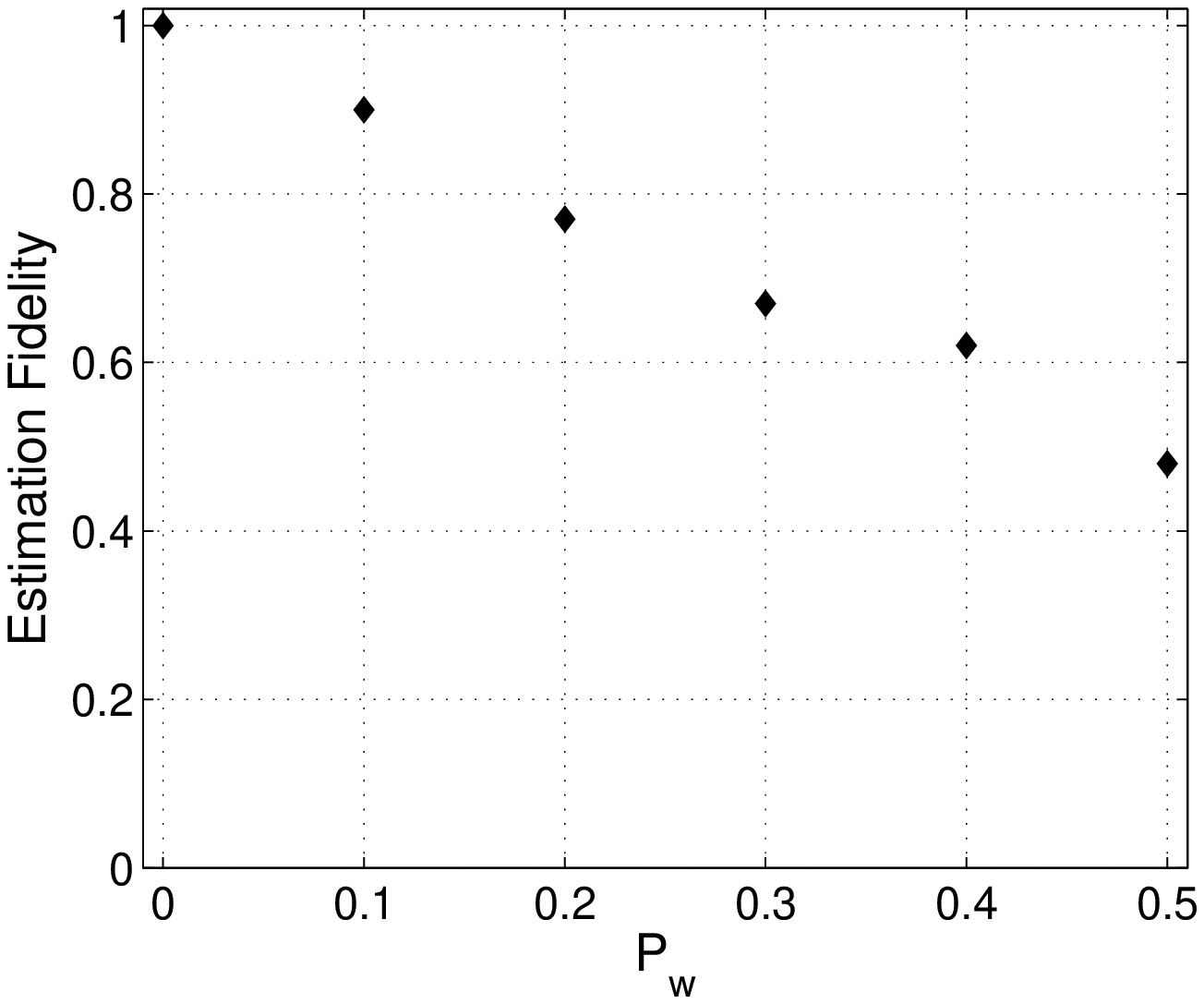}
 \caption{State estimation fidelity as a function of the probability, $P_w$, to make a wrong mapping during a QLS
measurement.\label{pwrongfig}} 
\end{figure}

 As discussed earlier imperfect state mapping during QLS measuremtents will be strongly affected by trap heating rates
which depend on the trap geometry. Rather than try to estimate the typical effect we therefore demonstrate what the
effect on estimation fidelity is as a function of different probabilities for an incorrect mapping.  This is plotted in
Fig.~\ref{pwrongfig} where we simulated a qubit undergoing Rabi oscillations in the absence of dephasing noise.  We used
$p_0=0.45$ and carried out ten unsharp measurement per oscillation of the $z$-component of the spin. After each
measurement we update our state estimate according to Eq.~(\ref{stateupdate}) with the correct, or incorrect,
measurement operator with probability $1-P_w$ or $P_w$, respectively.  As seen in Fig.~\ref{pwrongfig} imperfect
mapping at the level of about 10\% can
be tolerated at estimation fidelities around 90\%.

\subsubsection{Spontaneous Emission}

\begin{figure}
\includegraphics[scale=0.45]{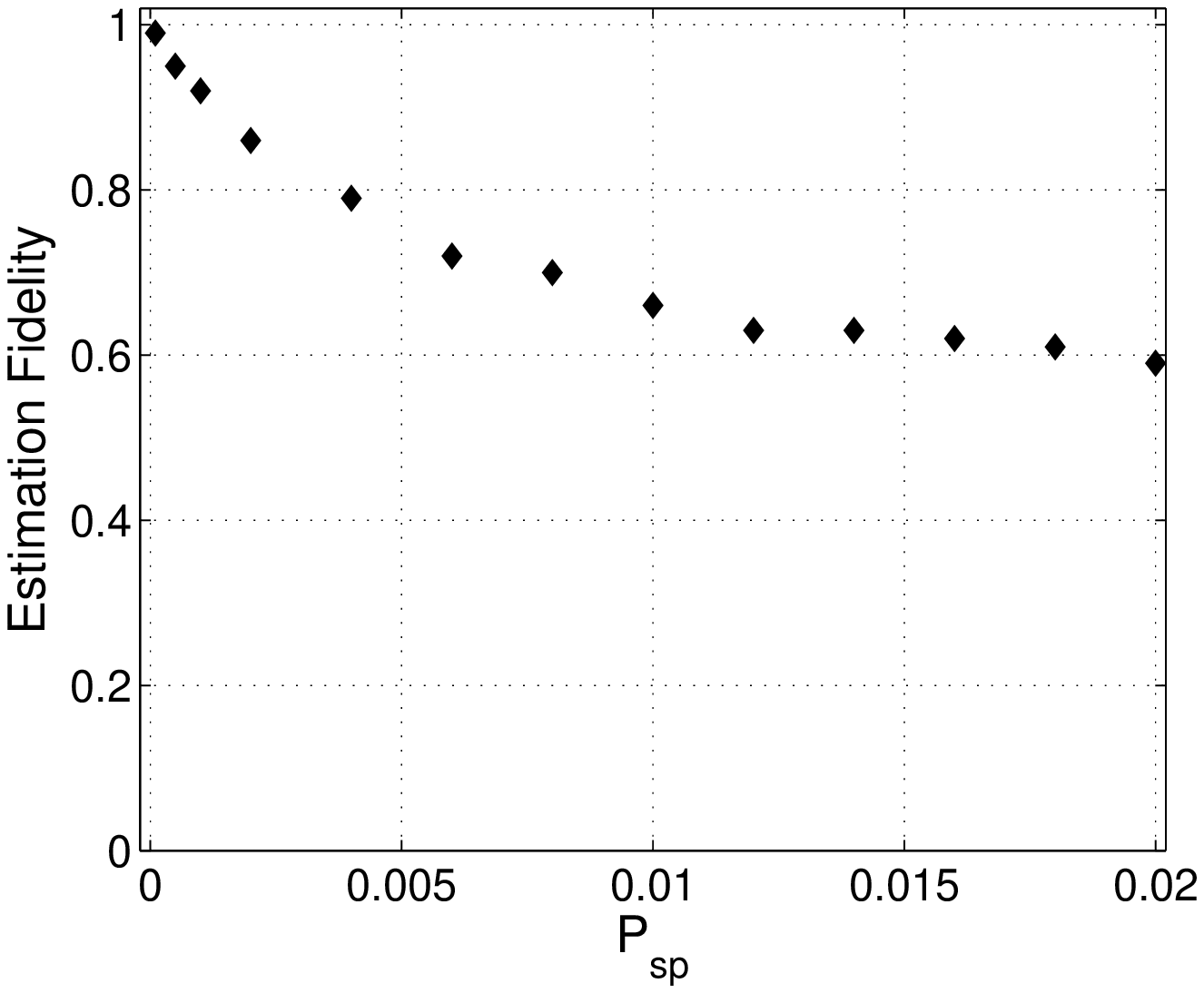}
 \caption{State estimation fidelity as a function of the probability, $P_w$, to make a wrong mapping during a QLS
measurement.\label{pspfig}} 
\end{figure}

In sections 4 A.1 and 4 B.1 we estimated that spontaneous scattering rates in typical experiments lead to a
probability of $P_{sp}\approx0.0005 - 0.001$ for a spontaneous scattering event per unsharp measurement.  We again
characterize
the estimation fidelity in a simulation of a two-level system undergoing Rabi oscillations, carrying out ten unsharp
measurements per oscillation of the $z$-component of the spin, and using $p_0=0.45$. During each
measurement we assume a probability of $P_{sp}$ for the qubit state to collapse into either of the upper or lower level
with equal probability.  The results are summarized in Fig.~\ref{pspfig}.  When compared to the the results of
Fig.~\ref{pwrongfig} we see that a fixed likelihood per measurement for a spontaneous scattering event, is significantly
more detrimental than the same likelihood for making an incorrect detection.  In particular, in this case
$P_{sp}\approx0.001$ yields an estimation fidelity of 90\%, while the same fidelity is achieved with $P_w\approx0.1$.

\subsection{State preparation using unsharp measurement}

As a second application we consider the state preparation scheme proposed by Ashab and
Nori \cite{Ashab2010}.  They demonstrated that it is possible to prepare arbitrary qubit states using
unsharp measurements along only a restricted set of orthogonal axes.  For preparing a state 
\begin{equation}
 |\psi\rangle = \sin{\frac{\theta_T}{2}}e^{+i\phi_T/2}|0\rangle+\cos{\frac{\theta_T}{2}}e^{-i\phi_T/2}|1\rangle,
\label{targetstate}
\end{equation}
and assuming the state is initially in $|\psi_0\rangle = (|0\rangle+|1\rangle)/\sqrt{2}$, they take the following
approach.\\
(1) Make a sequence of POVM measurements of $\hat\sigma_y$ until the qubit phase angle converges to $\phi_T$.\\
(2) Make a sequence of POVM measurements of $\hat\sigma_z$ until the polar angle converges to $\theta_T$.\\
During each of these two steps there is a roughly 50\% chance that the qubit wanders in the wrong angular direction. If
this
happens simply reset the qubit to the state $|\psi_0\rangle$ by making alternately strong, \textit{projective} 
measurements of $\hat\sigma_x$ and $\hat\sigma_y$ until the qubit ends up in $|\psi_0\rangle$.

We simulate this procedure in order to prepare the qubit state with angles $\theta_T=\pi/4$ and $\phi_T=\pi/2$. 
Choosing $p_0=0.15$ gives us fast convergence to the target state with a fidelity of
$F=|\langle\psi_0|\psi_T\rangle|^2=0.998$ within an average of 22 measurements. 
\begin{figure}
\includegraphics[scale=0.45]{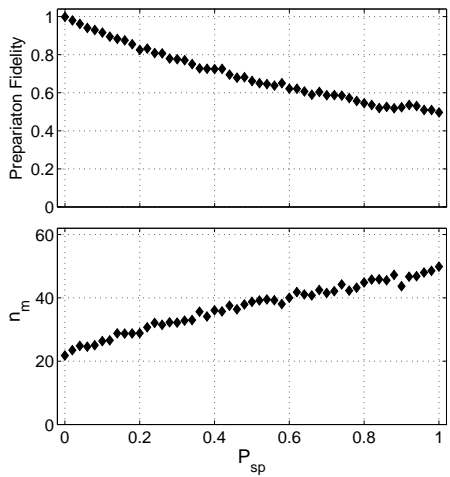}
 \caption{Performance of POVM measurement state preparation scheme in the presence of spontaneous emission. (a) State
preparation fidelity as a function of the probability per measurement for a spontaneous emission event to take place. 
(b) Number of measurements to convergence as a function of the probability per measurement for a spontaneous emission
event to take place.\label{prepFsp}} 
\end{figure}
Again we test the performance of the preparation scheme when spontaneous emission or imperfect state mapping is present
during the unsharp measurements. Figure \ref{prepFsp}(a) shows the decay of the state preparation fidelity as a function
of the probability per measurement, $P_{sp}$, for a spontaneous emission event to take place. The corresponding average
number of measurements to convergence is shown in Fig.~\ref{prepFsp}(b)
demonstrating that it requires more measurements to converge in the presence of stronger spontaneous scattering.
\begin{figure}
\includegraphics[scale=0.45]{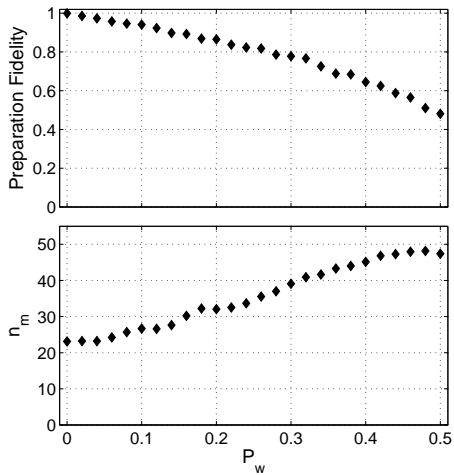}
 \caption{Performance of POVM measurement state preparation scheme in the presence of imperfect detection of the
auxiliary ion state. (a) State
preparation fidelity as a function of the probability per measurement for an incorrect detection to take place. 
(b) Number of measurements to convergence as a function of the probability per measurement for an incorrect detection to
take place.\label{prepFw}} 
\end{figure}
In Figs.~\ref{prepFw} (a) and (b) similar curves are plotted for different likelihoods of imperfect state mapping. 
We see that state preparation at fidelities above 90\% should be possible if the likelihood of spontaneous emission and
incorrect detection can be maintained below 10\% per measurement.

\section{Conclusion}
In conclusion, we have shown here how to realize a class of symmetric POVM measurements on the qubit levels of a
trapped ion.  Having analyzed the most common sources of error these measurements will be subject to, we
conclude that convincing proof-of-principle experiments of applications of unsharp measurements should be possible by
exploiting standard ion trap quantum control techniques. This work expands the set of tools available to ion trap
quantum control experimentalists and opens the door to exploring new applications such as real time quantum state
monitoring or quantum feedback using unsharp measurements.\\

\appendix

\section{Preparation of an arbitrary symmetric POVM}

We discuss how to implement a POVM measurement of the form defined by Eqs.~(\ref{PpPm}), (\ref{unsharp1}) and
(\ref{unsharp2}) for arbitrary $\hat r = \sin{\theta}\cos{\phi}\;\hat x  + \sin{\theta}\sin{\phi}\;\hat y
+\cos{\theta}\;\hat z$.  We only present an approach based on scheme I. Explicitly the two measurement operators are:\\
\begin{equation}
\hat M_0 =\frac{1}{2}\begin{pmatrix} 
 p_++p_-\cos{\theta} & p_-\sin{\theta}e^{-i\phi}\\
p_-\sin{\theta}e^{+i\phi} & p_+-p_-\cos{\theta}
\end{pmatrix},\label{M0}
\end{equation}
\newline
\newline
and
\newline
\begin{equation} 
\hat M_1 =\frac{1}{2}\begin{pmatrix} 
 p_+-p_-\cos{\theta} & -p_-\sin{\theta}e^{-i\phi}\\
-p_-\sin{\theta}e^{+i\phi} & p_++p_-\cos{\theta}
\end{pmatrix},\label{M1}
\end{equation}
\newline

\noindent where $p_\pm=\sqrt{p_0}\pm\sqrt{1-p_0}$. If the target/auxiliary state is initially:
\begin{equation}
 |\psi_0\rangle = \left(c_1|g\rangle+c_2|e\rangle\right)|1\rangle
\end{equation}
then to effect measurement operators Eqs.~(\ref{M0}), (\ref{M1}) we must prepare the target/auxiliary state
\begin{eqnarray}
 |\psi_0\rangle =&&\!\!\!\!
\left[\frac{c_1}{2}\left(p_+-p_-(\cos{\theta}-\sin{\theta}e^{-i\phi})\right)|g\rangle\right.\nonumber\\
    &&
\;\;\;\;\;\;\;\;\left.+\frac{c_2}{2}\left(p_++p_-(\cos{\theta}+\sin{\theta}e^{+i\phi})\right)|e\rangle\right]
|0\rangle\nonumber\\
&&\!\!\!\!+\left[\frac{c_1}{2}\left(p_++p_-(\cos{\theta}-\sin{\theta}e^{-i\phi})\right)|g\rangle\right.\nonumber\\
    &&
\;\;\;\;\;\;\;\;\left.+\frac{c_2}{2}\left(p_+-p_-(\cos{\theta}+\sin{\theta}e^{+i\phi})\right)|e\rangle\right]
|1\rangle.\nonumber\\
&&\label{M01state}
\end{eqnarray}
\noindent A projective measurement on the auxiliary state will then effect the POVM measurement on the target state.  
To simplify our discussion let's regroup the kets in Eq.~(\ref{M01state}) and write each coefficient as a single
complex number.
\begin{equation}
 |\psi\rangle = c_1\left(a_1|0\rangle+a_2|1\rangle\right)|g\rangle
+c_2\left(b_1|0\rangle+b_2|1\rangle\right)|e\rangle.\label{M01statesimp}
\end{equation}
Once the direction, $\hat r$, and the strength, $p_0$, of the POVM measurement have been chosen, the coefficients in
Eq.~(\ref{M01statesimp}) can be found by direct comparison with Eq.~(\ref{M01state}).  The desired state is prepared by
simply doing single qubit rotations on the transition $|g_t\rangle\rightarrow|r_t\rangle$ to have the same coefficients
$a_1$ and $a_2$,  as in Eq.~(\ref{M01statesimp}), during the first carrier pulse of the four pulse measurement
preparation sequence.  The RSB pulse then maps the auxiliary states to give the term multiplying
$|g_t\rangle$ in Eq.~(\ref{M01statesimp}). The procedure is then repeated on the $|e_t\rangle\rightarrow|r_t\rangle$
transition.


\begin{thebibliography}{22}
\expandafter\ifx\csname natexlab\endcsname\relax\def\natexlab#1{#1}\fi
\expandafter\ifx\csname bibnamefont\endcsname\relax
  \def\bibnamefont#1{#1}\fi
\expandafter\ifx\csname bibfnamefont\endcsname\relax
  \def\bibfnamefont#1{#1}\fi
\expandafter\ifx\csname citenamefont\endcsname\relax
  \def\citenamefont#1{#1}\fi
\expandafter\ifx\csname url\endcsname\relax
  \def\url#1{\texttt{#1}}\fi
\expandafter\ifx\csname urlprefix\endcsname\relax\def\urlprefix{URL }\fi
\providecommand{\bibinfo}[2]{#2}
\providecommand{\eprint}[2][]{\url{#2}}

\bibitem[{\citenamefont{Wineland et~al.}(2003)\citenamefont{Wineland, Barret,
  Britton, Chiaverini, DeMarco, Itano, Jelenkovic, Langer, Leibfried, Meyer
  et~al.}}]{Wineland2003}
\bibinfo{author}{\bibfnamefont{D.}~\bibnamefont{Wineland}},
  \bibinfo{author}{\bibfnamefont{M.}~\bibnamefont{Barret}},
  \bibinfo{author}{\bibfnamefont{J.}~\bibnamefont{Britton}},
  \bibinfo{author}{\bibfnamefont{J.}~\bibnamefont{Chiaverini}},
  \bibinfo{author}{\bibfnamefont{B.}~\bibnamefont{DeMarco}},
  \bibinfo{author}{\bibfnamefont{W.}~\bibnamefont{Itano}},
  \bibinfo{author}{\bibfnamefont{B.}~\bibnamefont{Jelenkovic}},
  \bibinfo{author}{\bibfnamefont{C.}~\bibnamefont{Langer}},
  \bibinfo{author}{\bibfnamefont{D.}~\bibnamefont{Leibfried}},
  \bibinfo{author}{\bibfnamefont{V.}~\bibnamefont{Meyer}},
  \bibnamefont{et~al.}, \bibinfo{journal}{Phil. Trans. R. Soc. Lond. A}
  \textbf{\bibinfo{volume}{361}}, \bibinfo{pages}{1349} (\bibinfo{year}{2003}).

\bibitem[{\citenamefont{Schmidt et~al.}(2005)\citenamefont{Schmidt, Rosenband,
  Langer, Itano, Bergquist, and Wineland}}]{Schmidt2005}
\bibinfo{author}{\bibfnamefont{P.}~\bibnamefont{Schmidt}},
  \bibinfo{author}{\bibfnamefont{T.}~\bibnamefont{Rosenband}},
  \bibinfo{author}{\bibfnamefont{C.}~\bibnamefont{Langer}},
  \bibinfo{author}{\bibfnamefont{W.}~\bibnamefont{Itano}},
  \bibinfo{author}{\bibfnamefont{J.}~\bibnamefont{Bergquist}},
  \bibnamefont{and} \bibinfo{author}{\bibfnamefont{D.}~\bibnamefont{Wineland}},
  \bibinfo{journal}{Science} \textbf{\bibinfo{volume}{309}},
  \bibinfo{pages}{2005} (\bibinfo{year}{2005}).

\bibitem[{\citenamefont{Hetet et~al.}(2011)\citenamefont{Hetet, Slodicka, and
  Blatt}}]{Hetet2011}
\bibinfo{author}{\bibfnamefont{G.}~\bibnamefont{Hetet}},
  \bibinfo{author}{\bibfnamefont{L.}~\bibnamefont{Slodicka}}, \bibnamefont{and}
  \bibinfo{author}{\bibfnamefont{R.}~\bibnamefont{Blatt}},
  \bibinfo{journal}{Phys. Rev. Lett.} \textbf{\bibinfo{volume}{107}},
  \bibinfo{pages}{133002} (\bibinfo{year}{2011}).

\bibitem[{\citenamefont{Blatt and Roos}(2012)}]{Blatt2012}
\bibinfo{author}{\bibfnamefont{R.}~\bibnamefont{Blatt}} \bibnamefont{and}
  \bibinfo{author}{\bibfnamefont{C.}~\bibnamefont{Roos}},
  \bibinfo{journal}{Nature Physics} \textbf{\bibinfo{volume}{8}},
  \bibinfo{pages}{277} (\bibinfo{year}{2012}).

\bibitem[{\citenamefont{{Nielsen M.A.} and Chuang}(2002)}]{NielsenChuang}
\bibinfo{author}{\bibnamefont{{Nielsen M.A.}}} \bibnamefont{and}
  \bibinfo{author}{\bibfnamefont{I.}~\bibnamefont{Chuang}},
  \emph{\bibinfo{title}{Quantum {C}omputation and {Q}uantum {I}nformation}}
  (\bibinfo{publisher}{Cambridge, UK}, \bibinfo{year}{2002}).

\bibitem[{\citenamefont{Sayrin et~al.}(2011)\citenamefont{Sayrin, Dotsenko,
  Zhou, Peaudecerf, Rybarczyk, Gleyzes, Rouchon, Mirrahimi, Amini, Brune
  et~al.}}]{Sayrin2011}
\bibinfo{author}{\bibfnamefont{C.}~\bibnamefont{Sayrin}},
  \bibinfo{author}{\bibfnamefont{I.}~\bibnamefont{Dotsenko}},
  \bibinfo{author}{\bibfnamefont{X.}~\bibnamefont{Zhou}},
  \bibinfo{author}{\bibfnamefont{B.}~\bibnamefont{Peaudecerf}},
  \bibinfo{author}{\bibfnamefont{T.}~\bibnamefont{Rybarczyk}},
  \bibinfo{author}{\bibfnamefont{S.}~\bibnamefont{Gleyzes}},
  \bibinfo{author}{\bibfnamefont{P.}~\bibnamefont{Rouchon}},
  \bibinfo{author}{\bibfnamefont{M.}~\bibnamefont{Mirrahimi}},
  \bibinfo{author}{\bibfnamefont{H.}~\bibnamefont{Amini}},
  \bibinfo{author}{\bibfnamefont{M.}~\bibnamefont{Brune}},
  \bibnamefont{et~al.}, \bibinfo{journal}{Nature}
  \textbf{\bibinfo{volume}{477}}, \bibinfo{pages}{73} (\bibinfo{year}{2011}).

\bibitem[{\citenamefont{Gillet et~al.}(2010)\citenamefont{Gillet, Dalton,
  Lanyon, Almeida, Barbieri, Pryde, O'Brien, Resch, Bartlet, and
  White}}]{Gillet2010}
\bibinfo{author}{\bibfnamefont{G.}~\bibnamefont{Gillet}},
  \bibinfo{author}{\bibfnamefont{R.}~\bibnamefont{Dalton}},
  \bibinfo{author}{\bibfnamefont{B.}~\bibnamefont{Lanyon}},
  \bibinfo{author}{\bibfnamefont{M.}~\bibnamefont{Almeida}},
  \bibinfo{author}{\bibfnamefont{M.}~\bibnamefont{Barbieri}},
  \bibinfo{author}{\bibfnamefont{G.}~\bibnamefont{Pryde}},
  \bibinfo{author}{\bibfnamefont{J.}~\bibnamefont{O'Brien}},
  \bibinfo{author}{\bibfnamefont{K.}~\bibnamefont{Resch}},
  \bibinfo{author}{\bibfnamefont{S.}~\bibnamefont{Bartlet}}, \bibnamefont{and}
  \bibinfo{author}{\bibfnamefont{A.}~\bibnamefont{White}},
  \bibinfo{journal}{Phys. Rev. Lett} \textbf{\bibinfo{volume}{104}},
  \bibinfo{pages}{080503} (\bibinfo{year}{2010}).

\bibitem[{\citenamefont{Konrad and Uys}(2012)}]{Konrad2012}
\bibinfo{author}{\bibfnamefont{T.}~\bibnamefont{Konrad}} \bibnamefont{and}
  \bibinfo{author}{\bibfnamefont{H.}~\bibnamefont{Uys}},
  \bibinfo{journal}{Phys. Rev. A} \textbf{\bibinfo{volume}{85}},
  \bibinfo{pages}{012102} (\bibinfo{year}{2012}).

\bibitem[{\citenamefont{Audretsch et~al.}(2007)\citenamefont{Audretsch, Klee,
  and Konrad}}]{Audretsch0107}
\bibinfo{author}{\bibfnamefont{J.}~\bibnamefont{Audretsch}},
  \bibinfo{author}{\bibfnamefont{F.}~\bibnamefont{Klee}}, \bibnamefont{and}
  \bibinfo{author}{\bibfnamefont{T.}~\bibnamefont{Konrad}},
  \bibinfo{journal}{Physics Letters A} \textbf{\bibinfo{volume}{361}},
  \bibinfo{pages}{212} (\bibinfo{year}{2007}).

\bibitem[{\citenamefont{Vijay et~al.}(2012)\citenamefont{Vijay, Macklin,
  Slichter, Weber, Murch, Naik, Korotkov, and Siddiqi}}]{VijayMacklin2012}
\bibinfo{author}{\bibfnamefont{R.}~\bibnamefont{Vijay}},
  \bibinfo{author}{\bibfnamefont{C.}~\bibnamefont{Macklin}},
  \bibinfo{author}{\bibfnamefont{D.}~\bibnamefont{Slichter}},
  \bibinfo{author}{\bibfnamefont{S.}~\bibnamefont{Weber}},
  \bibinfo{author}{\bibfnamefont{K.}~\bibnamefont{Murch}},
  \bibinfo{author}{\bibfnamefont{R.}~\bibnamefont{Naik}},
  \bibinfo{author}{\bibfnamefont{A.}~\bibnamefont{Korotkov}}, \bibnamefont{and}
  \bibinfo{author}{\bibfnamefont{I.}~\bibnamefont{Siddiqi}},
  \bibinfo{journal}{Nature} \textbf{\bibinfo{volume}{490}}, \bibinfo{pages}{77}
  (\bibinfo{year}{2012}).

\bibitem[{\citenamefont{Audretsch
  et~al.}(2002{\natexlab{a}})\citenamefont{Audretsch, Konrad, and
  Scherer}}]{Audretsch0202}
\bibinfo{author}{\bibfnamefont{J.}~\bibnamefont{Audretsch}},
  \bibinfo{author}{\bibfnamefont{T.}~\bibnamefont{Konrad}}, \bibnamefont{and}
  \bibinfo{author}{\bibfnamefont{A.}~\bibnamefont{Scherer}},
  \bibinfo{journal}{Phys. Rev. A} \textbf{\bibinfo{volume}{65}},
  \bibinfo{pages}{033814} (\bibinfo{year}{2002}{\natexlab{a}}).

\bibitem[{\citenamefont{Wineland et~al.}(1998)\citenamefont{Wineland, Monroe,
  Itano, Leibfried, King, and Meekhof}}]{Wineland1998}
\bibinfo{author}{\bibfnamefont{D.}~\bibnamefont{Wineland}},
  \bibinfo{author}{\bibfnamefont{C.}~\bibnamefont{Monroe}},
  \bibinfo{author}{\bibfnamefont{W.}~\bibnamefont{Itano}},
  \bibinfo{author}{\bibfnamefont{D.}~\bibnamefont{Leibfried}},
  \bibinfo{author}{\bibfnamefont{B.}~\bibnamefont{King}}, \bibnamefont{and}
  \bibinfo{author}{\bibfnamefont{D.}~\bibnamefont{Meekhof}},
  \bibinfo{journal}{J. Res. Natl. Inst. Stand. Technol.}
  \textbf{\bibinfo{volume}{103}}, \bibinfo{pages}{259} (\bibinfo{year}{1998}).

\bibitem[{\citenamefont{Kitagawa and Ueda}(1992)}]{Kitagawa1993}
\bibinfo{author}{\bibfnamefont{M.}~\bibnamefont{Kitagawa}} \bibnamefont{and}
  \bibinfo{author}{\bibfnamefont{M.}~\bibnamefont{Ueda}},
  \bibinfo{journal}{Phys. Rev. A} \textbf{\bibinfo{volume}{47}},
  \bibinfo{pages}{1993} (\bibinfo{year}{1992}).

\bibitem[{\citenamefont{Leibfried et~al.}(2003)\citenamefont{Leibfried,
  DeMarco, Meyer, Lucas, Barret, Britton, Itano, Jelenkovic, Langer, Rosenband
  et~al.}}]{Leibfried2003}
\bibinfo{author}{\bibfnamefont{D.}~\bibnamefont{Leibfried}},
  \bibinfo{author}{\bibfnamefont{B.}~\bibnamefont{DeMarco}},
  \bibinfo{author}{\bibfnamefont{V.}~\bibnamefont{Meyer}},
  \bibinfo{author}{\bibfnamefont{D.}~\bibnamefont{Lucas}},
  \bibinfo{author}{\bibfnamefont{M.}~\bibnamefont{Barret}},
  \bibinfo{author}{\bibfnamefont{J.}~\bibnamefont{Britton}},
  \bibinfo{author}{\bibfnamefont{W.}~\bibnamefont{Itano}},
  \bibinfo{author}{\bibfnamefont{B.}~\bibnamefont{Jelenkovic}},
  \bibinfo{author}{\bibfnamefont{C.}~\bibnamefont{Langer}},
  \bibinfo{author}{\bibfnamefont{T.}~\bibnamefont{Rosenband}},
  \bibnamefont{et~al.}, \bibinfo{journal}{Nature}
  \textbf{\bibinfo{volume}{422}}, \bibinfo{pages}{412} (\bibinfo{year}{2003}).

\bibitem[{\citenamefont{Uys et~al.}(2010)\citenamefont{Uys, Biercuk,
  VanDevender, Ospelkaus, Ozeri, and Bollinger}}]{Uys2010}
\bibinfo{author}{\bibfnamefont{H.}~\bibnamefont{Uys}},
  \bibinfo{author}{\bibfnamefont{M.}~\bibnamefont{Biercuk}},
  \bibinfo{author}{\bibfnamefont{A.}~\bibnamefont{VanDevender}},
  \bibinfo{author}{\bibfnamefont{C.}~\bibnamefont{Ospelkaus}},
  \bibinfo{author}{\bibfnamefont{R.}~\bibnamefont{Ozeri}}, \bibnamefont{and}
  \bibinfo{author}{\bibfnamefont{J.}~\bibnamefont{Bollinger}},
  \bibinfo{journal}{Phys. Rev. Lett.} \textbf{\bibinfo{volume}{105}},
  \bibinfo{pages}{200401} (\bibinfo{year}{2010}).

\bibitem[{\citenamefont{Deslauriers et~al.}(2006)\citenamefont{Deslauriers,
  Olmschenk, Stick, Hensinger, Sterk, and Monroe}}]{Deslauriers2006}
\bibinfo{author}{\bibfnamefont{L.}~\bibnamefont{Deslauriers}},
  \bibinfo{author}{\bibfnamefont{S.}~\bibnamefont{Olmschenk}},
  \bibinfo{author}{\bibfnamefont{D.}~\bibnamefont{Stick}},
  \bibinfo{author}{\bibfnamefont{W.}~\bibnamefont{Hensinger}},
  \bibinfo{author}{\bibfnamefont{J.}~\bibnamefont{Sterk}}, \bibnamefont{and}
  \bibinfo{author}{\bibfnamefont{C.}~\bibnamefont{Monroe}},
  \bibinfo{journal}{Phys. Rev. Lett.} \textbf{\bibinfo{volume}{97}},
  \bibinfo{pages}{103007} (\bibinfo{year}{2006}).

\bibitem[{\citenamefont{Ozeri et~al.}(2005)\citenamefont{Ozeri, Langer, Jost,
  DeMarco, Ben-Kish, Blakestad, Britton, Chiaverini, Itano, Hume
  et~al.}}]{Ozeri2005}
\bibinfo{author}{\bibfnamefont{R.}~\bibnamefont{Ozeri}},
  \bibinfo{author}{\bibfnamefont{C.}~\bibnamefont{Langer}},
  \bibinfo{author}{\bibfnamefont{J.}~\bibnamefont{Jost}},
  \bibinfo{author}{\bibfnamefont{B.}~\bibnamefont{DeMarco}},
  \bibinfo{author}{\bibfnamefont{A.}~\bibnamefont{Ben-Kish}},
  \bibinfo{author}{\bibfnamefont{B.}~\bibnamefont{Blakestad}},
  \bibinfo{author}{\bibfnamefont{J.}~\bibnamefont{Britton}},
  \bibinfo{author}{\bibfnamefont{J.}~\bibnamefont{Chiaverini}},
  \bibinfo{author}{\bibfnamefont{W.}~\bibnamefont{Itano}},
  \bibinfo{author}{\bibfnamefont{D.}~\bibnamefont{Hume}}, \bibnamefont{et~al.},
  \bibinfo{journal}{Phys. Rev. Lett.} \textbf{\bibinfo{volume}{95}},
  \bibinfo{pages}{030403} (\bibinfo{year}{2005}).

\bibitem[{\citenamefont{Ozeri}(2011)}]{Ozeri2011}
\bibinfo{author}{\bibfnamefont{R.}~\bibnamefont{Ozeri}},
  \bibinfo{journal}{Contemporary Physics} \textbf{\bibinfo{volume}{52}},
  \bibinfo{pages}{531} (\bibinfo{year}{2011}).

\bibitem[{\citenamefont{Ashab and Nori}(2010)}]{Ashab2010}
\bibinfo{author}{\bibfnamefont{S.}~\bibnamefont{Ashab}} \bibnamefont{and}
  \bibinfo{author}{\bibfnamefont{F.}~\bibnamefont{Nori}},
  \bibinfo{journal}{Phys. Rev. A} \textbf{\bibinfo{volume}{82}},
  \bibinfo{pages}{062103} (\bibinfo{year}{2010}).

\bibitem[{\citenamefont{Audretsch
  et~al.}(2002{\natexlab{b}})\citenamefont{Audretsch, Di\'osi, and
  Konrad}}]{AudretschDiosiKonrad02}
\bibinfo{author}{\bibfnamefont{J.}~\bibnamefont{Audretsch}},
  \bibinfo{author}{\bibfnamefont{L.}~\bibnamefont{Di\'osi}}, \bibnamefont{and}
  \bibinfo{author}{\bibfnamefont{T.}~\bibnamefont{Konrad}},
  \bibinfo{journal}{Phys. Rev. A} \textbf{\bibinfo{volume}{66}},
  \bibinfo{pages}{022310} (\bibinfo{year}{2002}{\natexlab{b}}),
  \bibinfo{note}{e-print quant-ph/0201078}.

\bibitem[{\citenamefont{Olmschenk et~al.}(2007)\citenamefont{Olmschenk, Younge,
  Moehring, Matsukevich, Maunz, and Monroe}}]{Olmschenk2007}
\bibinfo{author}{\bibfnamefont{S.}~\bibnamefont{Olmschenk}},
  \bibinfo{author}{\bibfnamefont{K.}~\bibnamefont{Younge}},
  \bibinfo{author}{\bibfnamefont{D.}~\bibnamefont{Moehring}},
  \bibinfo{author}{\bibfnamefont{D.}~\bibnamefont{Matsukevich}},
  \bibinfo{author}{\bibfnamefont{P.}~\bibnamefont{Maunz}}, \bibnamefont{and}
  \bibinfo{author}{\bibfnamefont{C.}~\bibnamefont{Monroe}},
  \bibinfo{journal}{Phys. Rev. A} \textbf{\bibinfo{volume}{76}},
  \bibinfo{pages}{052314} (\bibinfo{year}{2007}).

\bibitem[{\citenamefont{Uhrig}(2008)}]{Uhrig2008}
\bibinfo{author}{\bibfnamefont{G.}~\bibnamefont{Uhrig}}, \bibinfo{journal}{New.
  J. Phys.} \textbf{\bibinfo{volume}{10}}, \bibinfo{pages}{083024}
  (\bibinfo{year}{2008}).

\end{thebibliography}

\end{document}